\documentclass[aps,prb,twocolumn,showpacs,superscriptaddress,groupedaddress]{revtex4}  
\usepackage{graphicx}  
\usepackage{dcolumn}   
\usepackage{bm}        
\usepackage{amssymb}   
\usepackage{epsfig}
\usepackage{epstopdf}
\usepackage{gensymb}

\usepackage{color,soul}

\begin{document}

\widetext

\title{Vortex Dynamics and Phase Diagram in the Electron Doped Cuprate Superconductor Pr$_{0.87}$LaCe$_{0.13}$CuO$_4$}
\author{S. Salem-Sugui Jr.}
\affiliation{Instituto de Fisica, Universidade Federal do Rio de Janeiro, 21941-972 Rio de Janeiro, RJ, Brazil}
\author{P. V. Lopes}
\affiliation{Instituto de Fisica, Universidade Federal do Rio de Janeiro, 21941-972 Rio de Janeiro, RJ, Brazil}
\author{M. P. Kern}
\affiliation{Instituto de Fisica, Universidade Federal do Rio de Janeiro, 21941-972 Rio de Janeiro, RJ, Brazil}
\author{Shyam Sundar}
\affiliation{Department of Physics, Simon Fraser University, Burnaby, British Columbia, V5A 1S6, Canada}
\author{Zhaoyu Liu}
\affiliation{Department of Physics, University of Washington, Seattle, WA 98195-1560, USA}
\author{Shiliang Li}
\affiliation{Beijing National Laboratory for Condensed Matter Physics, Institute of Physics, Chinese Academy of Sciences, Beijing 100190, China}
\affiliation{School of Physical Sciences, University of Chinese Academy of Sciences, Beijing 100049, China}
\affiliation{Songshan Lake Materials Laboratory, Dongguan, Guangdong 523808, China}
\author{Huiqian Luo}
\affiliation{Beijing National Laboratory for Condensed Matter Physics, Institute of Physics, Chinese Academy of Sciences, Beijing 100190, China}
\affiliation{Songshan Lake Materials Laboratory, Dongguan, Guangdong 523808, China}
\author{L. Ghivelder}
\affiliation{Instituto de Fisica, Universidade Federal do Rio de Janeiro, 21941-972 Rio de Janeiro, RJ, Brazil}


\begin{abstract}

Second magnetization peak (SMP) in hole-doped cuprates and iron pnictide superconductors has been widely explored. However, similar feature in the family of electron-doped cuprates is not common. Here, we report the vortex dynamics study in the single crystal of an electron-doped cuprate Pr$_{0.87}$LaCe$_{0.13}$CuO$_4$ superconductor using dc magnetization measurements. A SMP feature in the isothermal $M(H)$ was observed for $H$$\parallel$$ab$-planes. On the other hand, no such feature was observed for $H$$\parallel$$c$-axis in the crystal. Using magnetic relaxation data, a detailed analysis of activation pinning energy via collective creep theory suggests an elastic to plastic creep crossover across the SMP. Moreover, for $H$$\parallel$$ab$, a peak in the temperature dependence of critical current density is also observed near 7 K, which is likely be related to a dimensional crossover (3D-2D) associated to the emergence of Josephson vortices at low temperatures. The anisotropy parameter obtained $\gamma$ $\approx$ 8-11 indicates the 3D nature of vortex lattice mainly for $H$$\parallel$$c$-axis. The $H$-$T$ phase diagrams for $H$$\parallel$$c$ and $H$$\parallel$$ab$ are presented.

\end{abstract}

\pacs{}
\maketitle

\section{Introduction}

Understanding the vortex dynamics in high-$T_c$ cuprates has been a matter of great interest because of its relevance for technological applications, such as powerful superconducting electromagnets used for fusion reactors and particle accelerators \cite{Melhem:2012}. However, in comparison to its hole doped counterpart the vortex dynamics in electron doped cuprates has not been widely explored. An experimental feature in the isothermal $M(H)$, known as "second magnetization peak" (SMP) or "peak effect" has been largely studied to probe its fundamental origin in the hole-doped cuprates \cite{Abulafia:1996, Yang:1993, Baruch:1999} and iron-pnictides \cite{Prozorov:2008, Senatore:2008, Yang:2008, SSugui:2010} as well as in conventional low-$T_c$ superconductors \cite{Lortz:2007, Das:2009}, but such feature is atypical in electron doped cuprates. 

Furthermore, most of the studies of the SMP in high-$T_c$ superconductors were conducted for $H$$\parallel$$c$-axis of the crystal as for instance in optimally doped YBaCuO \cite{Kwok:1994} and in Bi2212 \cite{Chikumoto:1992}. Also, the SMP has been observed for both $H$$\parallel$$c$-axis $H$$\parallel$$ab$-planes directions in Y-Ho-Ba-Cu-O cuprate \cite{feng1997}, but to the best of our knowledge a SMP exclusively for $H$$\parallel$$ab$-planes has not been reported in the literature. 
However, in the year 2007, an anomalous peak was observed in the ac-susceptibility measurements of an electron-doped Pr$_{0.88}$LaCe$_{0.12}$CuO$_4$ single crystal for $H$$\parallel$$ab$-planes \cite{Wang:2007}, and associated with a SMP in the case of isothermal $M(H)$ measurements. However, the discovery and the increased focus in the iron-pnictides superconductors around the same time \cite{Kamihara:2006} perhaps made the community lose interest in the Pr$_{1-x}$LaCe$_x$CuO (PLCCO) system. Yet some additional studies were made over the years \cite{Wilson:2006, Niestemski:2007, Kang:2007, Wilson:2007, Li:2008, Fujita:2008, Wang:2009, Adachi:2013, Horio:2016, Adachi:2016, Yamamoto:2016, Adachi:2017, Baqiya:2019}, addressing the effect of electron doped induced superconductivity, disappearance of the antiferromagnetic phase with the emergence of superconductivity, Cu spin correlations, annealing, and the symmetry of the order parameter, all compared to the well know case of the hole doped cuprates. Moreover, a rather complicated crystal growth process in electron-doped cuprates as compared to its hole-doped counterpart, and a challenging post-growth annealing treatment, are among the reasons behind the less intense research in electron doped cuprates. In contrast to the hole-doped cuprates, as-grown crystals with electron doping do not show superconductivity and require annealing in oxygen-poor environment to exhibit superconductivity through removal of small amounts of oxygen \cite{Lambacher:2010}. It should be mentioned that in hole doped cuprates, such as YBa$_2$Cu$_3$O$_x$, the oxygen content plays a major role in the superconducting properties, where a change in the oxygen content $x$ from 6.3 to 6.9 yield changes in the superconducting transition temperature from $\sim$ 18 K to 92 K with an increase in anisotropy as $x$ decreases \cite{veal,deak}.

As the present literature in the Pr$_{1-x}$LaCe$_x$CuO (PLCCO) system lacks studies of vortex dynamics (related to the origin of SMP) and the vortex phase diagram, we address this issue by performing a detailed study of the temperature ($T$), magnetic field ($H$) and the time dependence of the magnetization (magnetic relaxation) in a Pr$_{0.87}$LaCe$_{0.13}$CuO$_4$ single crystal. A SMP for $H$$\parallel$$ab$-plane is observed and explained in-terms of an elastic to plastic creep crossover via magnetic relaxation measurements. This may be understood by considering the elasticity of the vortex lattice to pin the vortex at lower free energy sites (disorder sites) by deforming it at the expense of increasing elastic energy of the vortex lattice. The minimum of the sum of the two energies (pinning energy and elastic energy) forms an equilibrium vortex lattice configuration. At higher magnetic fields pinning gets stronger when the pinning energy at disordered sites overcomes the elastic energy of the vortex lattice and consequently yield a SMP in the isothermal magnetization $M(H)$. Below the SMP in the elastic creep regime the activation energy for the vortex creep increases with increasing magnetic field. On the other hand, above the SMP, when vortex creep is influenced by a plastic deformation of the vortex lattice (plastic creep, where large pieces of the vortex lattice slide \cite{Abulafia:1996}) the activation energy for the vortex creep decreases with increasing magnetic field. Moreover, the pinning behavior is explored through the temperature dependence of the critical current density where a crossover from 3D Abrikosov to a 2D Josephson vortices is observed at $\sim$ 7 K. Detailed vortex phase diagrams are also constructed for $H$$\parallel$$c$-axis and $H$$\parallel$$ab$-plane directions.

\section{Experimental Details}

High-quality single crystals of PLCCO (x=0.13) were grown by the traveling-solvent floating zone method using a mirror furnace (Cyberstar, France) \cite{Li:2008, Lambacher:2010}. The crystals have a tetragonal unit cell (space group I4/mmm) with the lattice parameters $a$ = $b$ = 3.981 \AA, $c$ = 12.27 \AA ~ which consists of two formula units. 
In the present study, we used a single crystal of mass $\sim$ 2.92 mg and the dimensions $\sim$ 1.3$\times$1.0$\times$0.2 mm$^3$. Magnetization measurements were performed for $H$$\parallel$$c$-axis and $H$$\parallel$$ab$-planes using a vibrating sample magnetometer (VSM) manufactured by the Quantum Design, with a quartz rod used for sample mounting. Data for temperature and magnetic field dependence of magnetization, $M$($T$) and $M$($H$) respectively, were collected in zero field cooled (ZFC) and field cooled (FC) states. In the ZFC state, the sample was cooled from above $T_c$ in zero applied magnetic field to the desired temperature below $T_c$ and the data was collected during warming, followed by collection of data during cooling in the FC state. For $H$$\parallel$$ab$-planes magnetic relaxation, $M(time)$, curves were also obtained for a span time of 100 minutes to explore the behavior of the SMP in the sample. 

\section{Results and Discussion}

\begin{figure}
\includegraphics[scale=0.42]{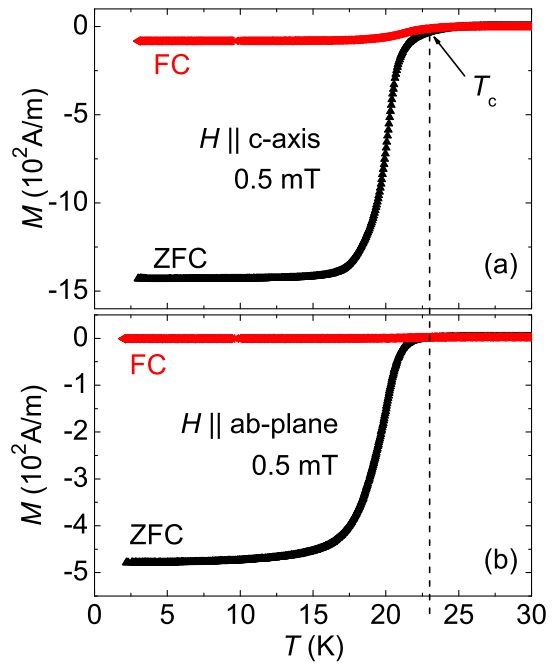}
\caption{\label{fig:Tc} Temperature dependence of magnetization measured at $H$ = 0.5 mT for {\bf(a)} $H$$\parallel$$c$-axis. {\bf(b)} $H$$\parallel$$ab$-planes. The splitting between the ZFC and FC curves occurs just below $T_c$ $\sim$ 23 K evidencing the strong vortex pinning in the sample }
\end{figure}

Figures \ref{fig:Tc}(a) and \ref{fig:Tc}(b) show $M$($T$) curves for $H$$\parallel$$c$-axis and $H$$\parallel$$ab$-planes respectively for an applied magnetic field $H$ = 0.5 mT. A significant drop in $M$($T$) has been observed at $\sim$ 23 K (see the dashed line) for both crystal directions which is defined as the onset of  the superconducting transition temperature, $T_c$ of the sample.  Similar compounds with a Ce content of 0.13 \cite{Baqiya:2019} and 0.12 \cite{Wang:2007, Roeser:2009}, have reported transition temperatures in the range 22-24 K, in good agreement with the present work. 

\begin{figure*}
\includegraphics[scale=0.48]{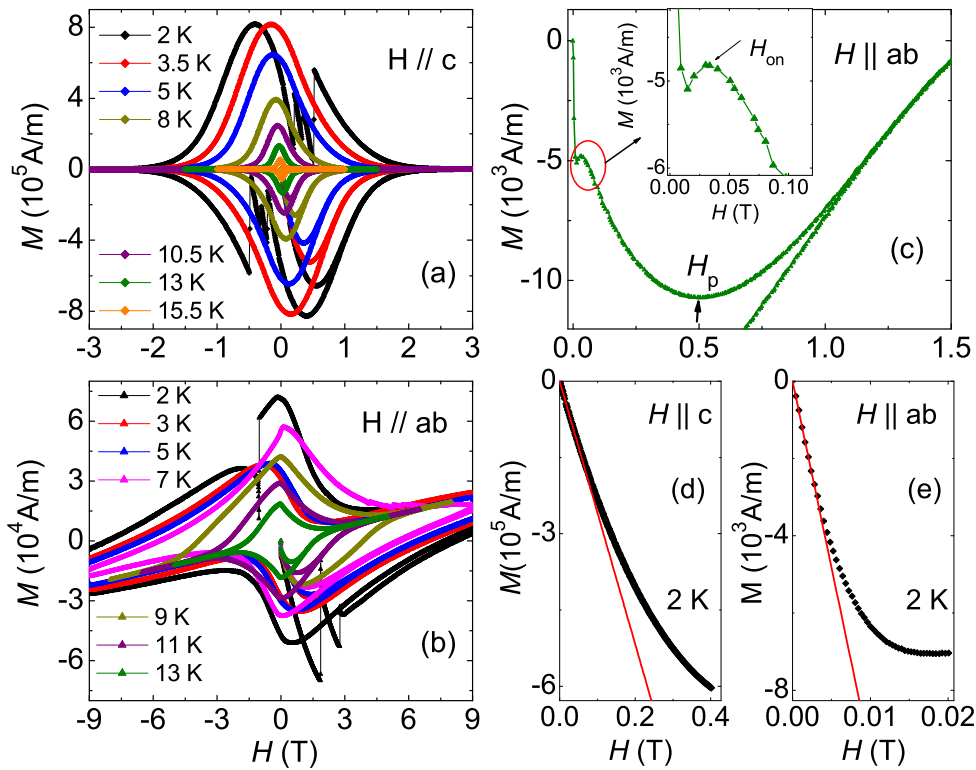}
\caption{\label{fig:MH} Isothermal magnetic field dependence of magnetization, $M$($H$), for selected temperatures, {\bf(a)} $H$$\parallel$c-axis, {\bf(b)} $H$$\parallel$$ab$-planes. {\bf(c)} Clear signature of SMP in $M$($H$) measured at $T$ = 13 K for $H$$\parallel$$ab$-planes, {\bf(d)} and {\bf(e)} show initial $M$($H$) curves at $T$ = 2K for both field directions, evidencing the Meissner region. Inset shows the enlarged view of the region near $H_{on}$.}
\end{figure*}

Figures \ref{fig:MH}(a) and \ref{fig:MH}(b) show selected isothermal $M$($H$) hysteresis curves for $H$$\parallel$$c$-axis and $H$$\parallel$$ab$-planes respectively. A SMP feature is observed for $H$$\parallel$$ab$-planes in the initial branch of $M$($H$) curves measured for $T$ $\le$ 15 K. Figure \ref{fig:MH}(c) exhibits the observed SMP for $T$ = 13 K and the associated peak field ($H_p$) and the onset field ($H_{on}$). It is to note that the observed $H_{on}$ values are more than an order of magnitude smaller than the $H_p$ values which obscure its appearance in the isothermal $M$($H$) curves shown in Fig. \ref{fig:MH}(b). Interestingly, we also observed that the clear SMP feature occurs only in the initial branch of the $M$($H$) curves (1st quadrant) and no such feature was found in other magnetic field branches. Nevertheless, it is possible to see, mainly for the $M$($H$) curves below 9 K that the decreasing field branch (2$^{nd}$ quadrant) shows a downward curvature for high fields which is followed by an upward curvature at intermediate fields and again becoming downward as $H$ approaches zero. Moreover, such curvature changes were also noticed in the following quadrants and seems to be reminiscent of the SMP observed in the 1$^{st}$ quadrant of $M$($H$) curves. We observed that $H_p$ increases as temperature decreases, while the same effect with temperature is not followed by the onset field $H_{on}$. Moreover, signatures of flux jumps are observed in the isothermal $M(H)$ measured at 2 K for both magnetic field directions.  Such magnetic instabilities or flux jumps in isothermal $M$($H$) usually occur at low temperatures where most of the materials have poor thermal conduction (or low heat capacity). Therefore, during flux flow, when the flux lines are forced to move against the large or irregular pinning force, the energy dissipation may locally increase the temperature of the sample. The increased temperature reduce the $J_c$ which requires the intake of more flux inside the sample to restore the critical state and may produce an avalanche of flux jumps \cite{Mints:1981}. Such magnetic instabilities have been observed in various low-$T_c$ \cite{Sundar:2015, Lee:2015} as well as in high-$T_c$ \cite{Li:2013, Garber:1993} superconductors. Figures \ref{fig:MH}(d) and (e) show the initial branch of $M$($H)$ curves at $T$ = 2 K for both field directions, allowing us to estimate the values of the lower critical field, $H_{c1}$(2K) $\approx$ 23 mT for $H$$\parallel$$c$-axis and $H_{c1}$(2K) $\approx$ 2.4 mT for $H$$\parallel$$ab$-planes. It should be mentioned that with the exception of Figs \ref{fig:MH}(d) and (e), all analysis performed in this work were conducted by using magnetic fields well above the Meissner region.

\begin{figure}
\includegraphics[scale=0.4]{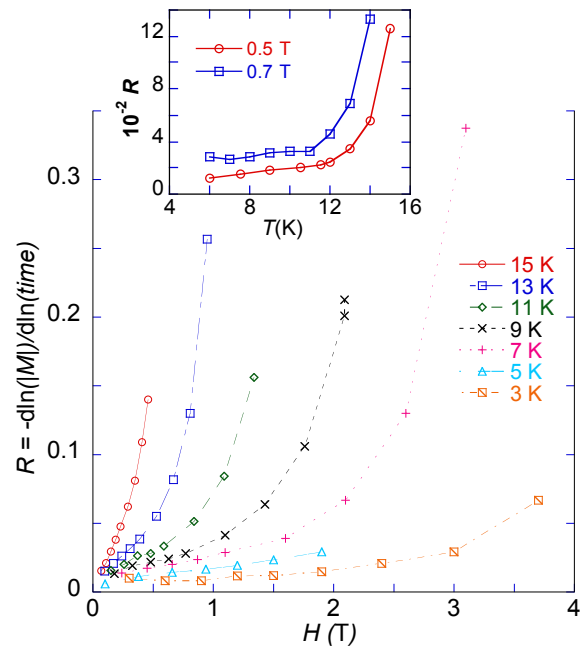}
\caption{\label{fig:RvsH} The relaxation rate $R$ = -dln$|M|$/dln$(time)$ plotted against $H$ as obtained for fields below and above the SMP for different isothermal $M(H)$. The inset shows $R$ plotted as a function of temperature as obtained for $H$ = 0.5 and 0.7 T.}
\end{figure}

The existence of a SMP in $M$($H$) curves with $H$$\parallel$$ab$-planes and not for $H$$\parallel$$c$-axis is unique and to the best of our knowledge has not been observed in any other material. To study this effect in more detail we have obtained several magnetic relaxation curves at different temperatures and magnetic fields for $H$$\parallel$$ab$-planes. The $H$ values for magnetic relaxation measurements were selected below and above the SMP in order to explore the behaviour across SMP. All $M(time)$ curves showed a linear logarithmic time dependence allowing to obtain the relaxation rate $R$ = -dln$|M|$/dln$(time)$. The main panel in Fig. \ref{fig:RvsH} shows the resulting $R$ vs. $H$ and the $R$ vs. $T$ dependence is depicted in the inset. The relaxation rate increases monotonically with the magnetic field and temperature and no apparent change in $R$ is observed which could be associated to the SMP in $M$($H$) curves for $H$$\parallel$$ab$-planes. Therefore, we used the Maley’s approach \cite{Maley:1990} in order to perform an analysis of the pinning activation energy behavior in $M$($H$) curves for magnetic fields across the peak field $H_p$.

\begin{figure}
\includegraphics[scale=0.4]{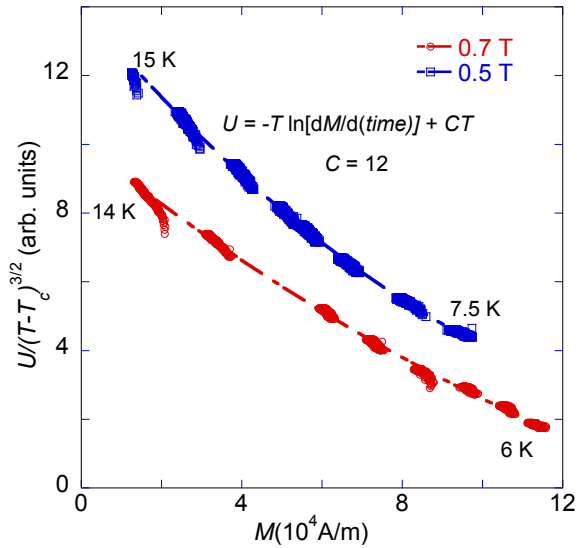}
\caption{\label{fig:Umaley} $U(M)$/$(T-T_c$)$^{3/2}$ plotted against $M$ with the final curves exhibiting a smooth behavior. Dashed lines are only a guide to the eyes.}
\end{figure}

The Maley’s approach establishes that the activation energy $U(M)$ = -$T$ln[d$M(time)$/d$(time)$] + $C$$T$ where $M$($time$) is the magnetic relaxation curve and $C$ is an intrinsic constant \cite{Maley:1990}, is a smooth function of $M$-$M_{eq}$ for a given magnetic field, where $M_{eq}$ is the equilibrium magnetization. In this approach the smooth curve of an isofield $U$($M$) curve can be obtained adjusting the value of the constant $C$. It was shown later \cite{McHenry:1991} that not always a single value of $C$ allows us to obtain a smooth curve of $U$($M$). This may require the scaling of $U$($M$) by a function that express how the coherence length correlates with temperature. Following this approach we used the scaling function $g(T$)=($T$-$T_c$)$^{3/2}$ as in Ref.\cite{McHenry:1991}. Figure \ref{fig:Umaley} shows the scaled $U(M)$ vs. $M$ smooth curves obtained at different temperatures for $H$ = 0.7 and 0.5 T with $C$ = 12. This value of $C$ allows us to calculate and study the isofield activation energy curves from magnetic relaxation data obtained for fields below and above the SMP in $M(H)$ curves.

\begin{figure}
\includegraphics[scale=0.45]{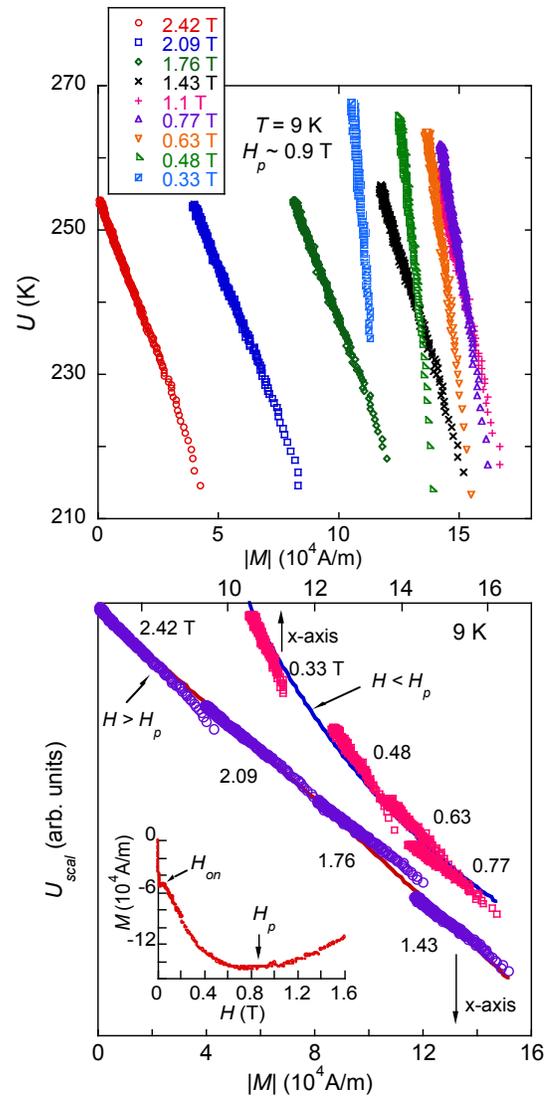}
\caption{\label{fig:figUscal} {\bf(a)} Activation energy $U(M)$ plotted against $M$. Values of $U(M)$ were obtained for fields below and above the SMP appearing in the $M(H)$ at 9 K. {\bf(b)} The scaled activation energy $U_{scal}$ is plotted against $M$, as obtained from fields below (empty squares) and above (empty circles) $H_p$. The x-axis of each smooth curve is shown by arrows. The inset shows details of the SMP occurring in the increasing field branch of the respective $M(H)$ curve.}
\end{figure}

 Figure \ref{fig:figUscal}(a) shows $U$($M$) obtained for $T$ = 9 K with $C$ = 12, where it is possible to see a change in the behavior of $U(M)$ for fields across $H_p$. We mention that the behavior of $U$($M$) with $M$ shown in Fig. \ref{fig:figUscal}(a) perfectly matches with the behavior of $U$($M$) observed in the SMP of YBaCuO \cite{Abulafia:1996} and suggests that the SMP might be associated to an elastic to plastic creep crossover. Such creep crossover across $H_p$ may be clearly observed by invoking a theory of collective flux creep \cite{Feigel'man:1989}. In this theory, the activation energy is defined as, $U$($B$,$J$) = $B^{\nu}$$J^{-\mu}$ $\approx$ $H^{\nu}$$M^{-\mu}$, where the exponents $\nu$ and $\mu$ depend on the speciﬁc flux-creep. The $U$($M$) curves in Fig. \ref{fig:figUscal}(a) were scaled with $H^{\nu}$ and the resulting smooth $U_{scal}$ curves of $U(M)$/$H^{0.5}$ and $U(M)$/$H^{-0.9}$ are shown in Fig. \ref{fig:figUscal}(b) for fields below and above $H_p$ respectively. The observed scaling behaviour across $H_p$ demonstrates the elastic to plastic creep crossover as the origin of the SMP for $H$$\parallel$ab-planes. One should note that there are two x-axis in Fig. \ref{fig:figUscal}(b). Similar exponents of $H$ for both elastic and plastic pinning have been observed in different studies on iron-pnictide superconductors \cite{Sundar1:2017, Sundar2:2017, Sundar:2019}. As there is no SMP for $H$$\parallel$$c$-axis the major pinning contribution for producing the SMP for $H$$\parallel$$ab$-planes is the pinning associated with the vortices between adjacent layers of Cu-O planes, which is known as intrinsic pinning \cite{Tachiki:1989}.

\begin{figure*}
\includegraphics[scale=0.45]{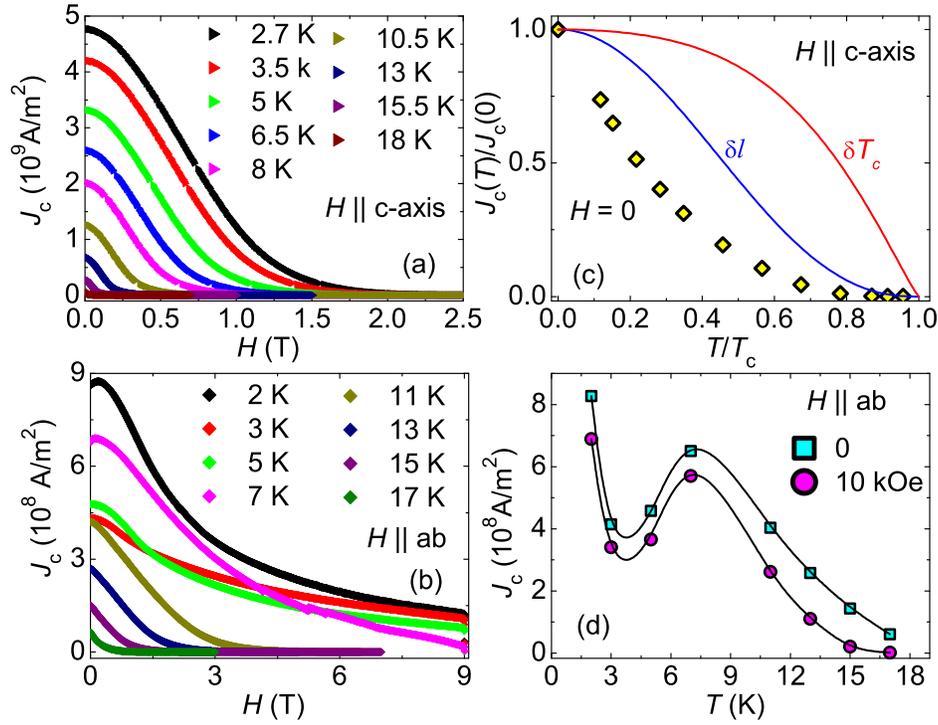}
\caption{\label{fig:Jc} Magnetic field dependence of critical current density at some selected temperatures {\bf(a)} for $H$$\parallel$$c$-axis. {\bf(b)} for $H$$\parallel$$ab$-planes. {\bf(c)} Normalized critical current density, $J_c(T)$/$J_c(0)$, as a function of reduced temperature, $t$ = $T/T_c$. Solid lines represent the behavior of $\delta$$T_c$ and $\delta$$l$ pinning models \cite{Griessen:1994} discussed in the text. {\bf(d)} Temperature dependence of critical current density for $H$$\parallel$$ab$-planes.}
\end{figure*}

Figure \ref{fig:Jc}(a) and \ref{fig:Jc}(b) show the magnetic field dependence of the critical current density, $J_c$($H$), at different temperatures for $H$$\parallel$ $c$-axis and $H$$\parallel$$ab$-planes respectively. $J_c$ was obtained from the isothermal $M$($H$) curves using the Bean's model \cite{Bean:1964} expression, $J_c$ = $\Delta$$M$/$[a_1$(1-$a_1$/3$a_2$)] \cite{Sundar1:2017}, where 2$a_2$ $>$ 2$a_1$, are the crystal dimensions of the plane perpendicular to $H$, and $\Delta$$M$(A/m) is the difference between the magnetic field decreasing and increasing branches of the $M$($H$) curve. The critical current density at all temperatures in zero field limit, $J_c(0)$, for $H$$\parallel$$c$-axis exceeds by about an order of magnitude when compared to $J_c(0)$ for $H$$\parallel$$ab$-planes, which is expected due to the strong pinning in the Abrikosov vortex-lattice along c-axis than in the intrinsic pinning (possibly of Josephson vortices at low temperatures) in the $ab$-planes. The clear signature of SMP appears only in the initial branch of the isothermal $M$($H$) curves and exhibits a change of slope in the following respective $M$($H$) branches (as also discussed earlier in text). Therefore, the usual peak feature in the $J_c(H)$ curves at $H_p$ \cite{Sundar2:2017} was not observed, instead a peak very close to $H$ = 0 is present in $J_c(H)$ for $T$ $<$ 7 K. Moreover, the $J_c$($H$) curves for $H$$\parallel$ $c$-axis show the usual behavior, where $J_c$ reduces as we increase $T$. On the other hand, $J_c$($H$) in the low field side for $H$$\parallel$$ab$-planes at $T$ = 5 K, 7 K is greater than its value at 3 K. This behavior is not usual and may indicate some change in pinning or vortex-lattice dimensional crossover across these temperatures in ab-planes. However, it clearly suggests that the pinning in the two crystallographic directions is quite different in nature. 

Therefore, it would be interesting to explore the temperature dependence of the critical current density in both crystal directions. A plot of the normalized critical current density, $J_c(T)$/$J_c(0)$ as a function of reduced temperature $t$ (=$T$/$T_c$) allows us to compare whether the dominant pinning in the sample is due to the variation in the charge carrier mean free path, called $\delta$$L$ pinning (commonly observed in pnictides \cite{Sundar:2019,gennep:2020}), where the respective expression is given by $J_c(T)$/$J_c(0)$ = $(1+t^2)$$^{-1/2}$$(1-t^2)$$^{5/2}$, or due to the variation of the superconducting transition temperature $T_c$, called $\delta$$T_c$ pinning, with $J_c(T)$/$J_c(0)$ = $(1-t^2)$$^{7/6}$$(1-t^2)$$^{5/6}$ \cite{Griessen:1994}. In Fig. \ref{fig:Jc}(c) it is observed that the $J_c(T)$/$J_c(0)$ data obtained for $H$=0 is not consistent with the model 
developed by Griessen et al for conventional pinning mechanisms \cite{Griessen:1994}. A rigorous analysis based on the model developed in Ref. \cite{Hosseinzadeh:2019} is required to better explain the dominant pinning mechanism in the sample. For $H$$\parallel$$ab$-planes, a peak in $J_c$ vs $T$ is observed which is shown for $H$ = 0 and 1 T in Fig. \ref{fig:Jc}(d). It suggests an abrupt change in the type of pinning or a change in the vortex-lattice occurs as temperature drops below 7 K. Since no marked changes were observed in the magnetic relaxation curves shown in Fig. \ref{fig:RvsH}, it is unlikely that a change in the type of pinning occurs below 7 K. Most likely the drop in $J_c$($T$) is related to a dimensional crossover (3$D$-2$D$) where 3D cylindrical vortices change to 2D Josephson vortices below 7 K. This is possible since the coherence length decreases with temperature with the subsequent emergence of Josephson vortices in a very anisotropic lattice \cite{Hu:1998}. The transition of an Abrikosov vortex lattice to a very anisotropic Josephson vortex lattice  \cite{Hu:1998} is expected to be followed by a change in the magnetic flux inside the sample. Also, Josephson vortices are more weakly pinned than Abrikosov vortices \cite{Fehresnbacher:1992}. This can be realized as Abrikosov vortices have two length scales, the London penetration depth $\lambda$ and the core which is the size of the coherence length, where the superconducting order parameter vanishes \cite{Fehresnbacher:1992}. On the other hand, Josephson vortices do not have a normal core \cite{Fehresnbacher:1992}, and are characterized by a single length scale given by $\lambda$$_J$ = ($c$$\phi_0$/16$\pi$$^2$$\lambda$$j_s$)$^{1/2}$ where $\phi_0$ is the quantum flux, $c$ is the light velocity, $\lambda$ is the London penetration depth and $j_s$ is the Josephson critical curent density, where $j_s$ is much smaller than both the critical current density $J_c$ and the depairing current density $j_d$ \cite{Gurevich:1992}. Usually $\lambda$$_J$ is much larger than $\lambda$, and Josephson vortices are more weakly pinned than Abrikosov vortices \cite{Gurevich:1992}. The existence of such a dimensional crossover is well known to occur in layered high-$T_c$ superconductors \cite{Baruch:2001}. A similar peak in $J_c$ vs $T$ for $H$$\parallel$$ab$ has also been observed in the pnictide superconductor SmFeAs(O,F) with $T_c$ $\sim$ 48-50 K and $\gamma$ $\sim$ 4-6 \cite{Moll:2013}. It was argued that such peak is a consequence of a transition from well-pinned, slow moving Abrikosov vortices at high temperatures to weakly pinned, fast ﬂowing Josephson vortices at low temperatures \cite{Moll:2013}. In addition, recently in Ref. \cite{Egilmez:2019}, the authors also observed a similar peak at intermediate temperature in $J_c$ vs. $T$ in YBa$_2$Cu$_4$O$_8$ crystals, however for $H$ tilted away from the $ab$ plane and the same peak is absent when $H$ is aligned along $c$- or $b$-axis of the crystal lattice. Such peak in $J_c$ vs. $T$ in YBa$_2$Cu$_4$O$_8$ crystal was explained in terms of thermally activated vortex-lattice instability, which splits the lattice and leads to vortex segments aligned along c-axis \cite{Egilmez:2019}. It is worth noticing that in the present magnetization measurements $H$ is applied along the ab-plane (i.e close to zero tilt angle between $H$ and ab-plane). However, we do not completely discard the possibility of a small misalignment ($<$ 1$^{\circ}$) in our measurements. Unfortunately, we do not have a built-in facility to control angle misalignment to confirm this argument. Regarding the sample quality, the surface attached to the sample holder was microscopically smooth with no indication of faulting  planes within the microscopic resolution. Despite not related to our results, we mention that a continuous vortex lattice transformation was observed in FeSe for magnetic fields going from 1 to 6 T for $H$$\parallel$ab \cite{putilov:2019}.

\begin{figure}
\includegraphics[scale=0.45]{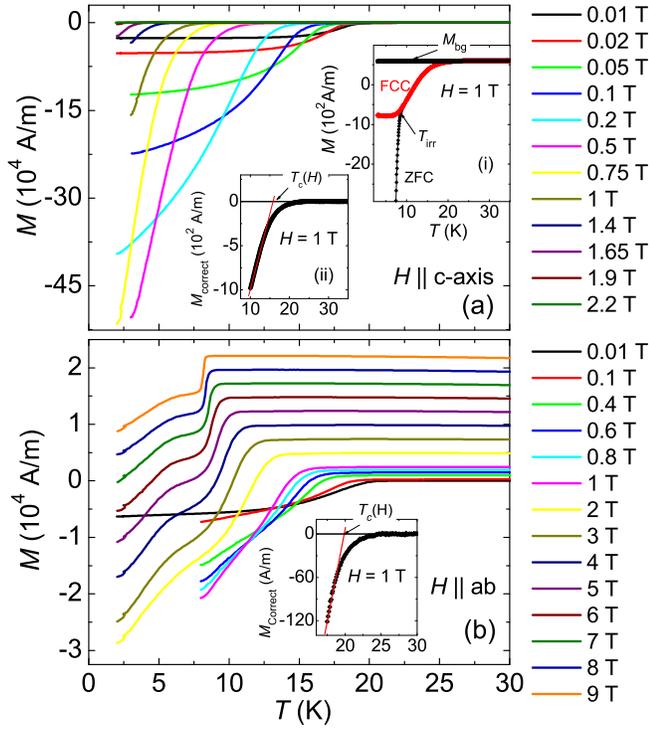}
\caption{\label{fig:MT} Isofield temperature dependence of ZFC magnetization measured at different applied magnetic field {\bf(a)} for $H$$\parallel$$c$-axis. {\bf(b)} for $H$$\parallel$$ab$-planes.}
\end{figure}

\begin{figure}
\includegraphics[scale=0.45]{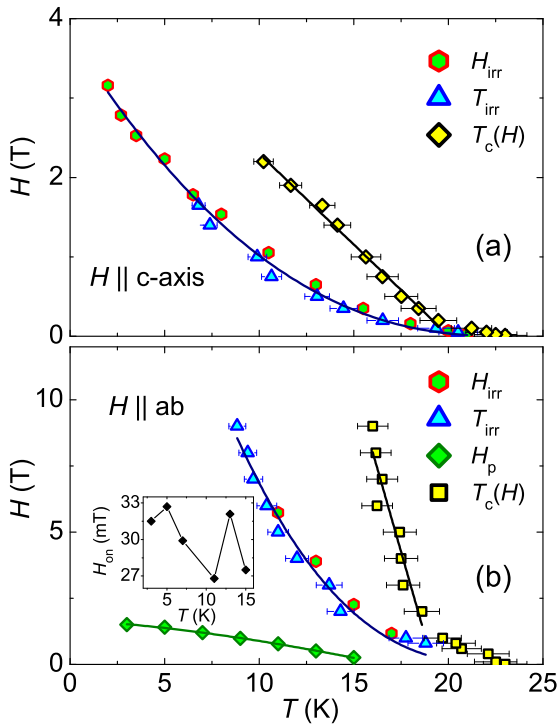}
\caption{\label{fig:HT} $H$-$T$ phase diagram {\bf(a)} for $H$$\parallel$$c$-axis. {\bf(b)} for $H$$\parallel$$ab$- planes. Solid lines show the fitted curves (see text for details). Inset shows the temperature variation of $H_{on}$.}
\end{figure}

Figures \ref{fig:MT}(a) and \ref{fig:MT}(b) show isofield $M(T)$ curves obtained for $H$$\parallel$ $c$-axis and $H$$\parallel$ $ab$-planes respectively. The well resolved split between the ZFC and FC curves in each $M(T)$ allowed to obtain the irreversible temperature, $T_{irr}$ as a function of magnetic field (see inset (i) in Fig. \ref{fig:MT}(a)). In addition, we also obtained the mean field transition temperature, $T_c(H)$ for each measured isofield $M(T)$ by the crossing point of the extended reversible region with the extension of the normal region \cite{Welp:1989}. The insets of Fig. \ref{fig:MT}(a) and \ref{fig:MT}(b) show this method applied to selected $M(T)$ curves obtained after background correction. The kinks appearing on the $M$($T$) curves for $H$$\parallel$ $ab$ at $\sim$ 7 K are likely to be associated to the decrease of $J_c$ observed below 7 K in Fig. \ref{fig:Jc}(d). As discussed above the possible emergence of Josephson vortices below 7K is expected to be followed by a change in the magnetic flux inside the sample. An $H$-$T$ phase diagram is constructed using the obtained $T_c(H)$ and $T_{irr}$ from each isofield $M(T)$ along with the characteristic fields $H_{irr}$, $H_p$ and $H_{on}$ from isothermal $M(H)$ measurements. A criteria, $J_c$ = 20x10$^4$ A/m$^2$ (for $H$$\parallel$ $c$) and $J_c$ = 50x10$^4$ A/m$^2$ (for $H$$\parallel$ $ab$) is used to consistently obtained the irreversibility field, $H_{irr}$ from each isothermal $M(H)$. The resulting $H$-$T$ phase diagrams for both crystal directions are shown in Fig. \ref{fig:HT}.

The upper critical field as a function of temperature, $H_{c2}(T)$ (or $T_c(H)$-line), for both crystal directions show smaller slope at low magnetic fields and a relatively larger slope at higher fields. Linear fit to the $H_{c2}(T)$ gives slopes of -0.23 T/K and -2.51 T/K for $H$$\parallel$$c$-axis and $H$$\parallel$$ab$-planes respectively. The anisotropy parameter, $\gamma_{H_{c2}}$ = (d$H_{c2}$/d$T$)$_{ab}$/(d$H_{c2}$/d$T$)$_c$ is estimated as $\sim$ 11 suggesting the 3D character of the vortex lattice in Pr$_{0.87}$LaCe$_{0.13}$CuO$_4$ crystal mainly for $H$$\parallel$$c$-axis. The orbital-limit of the upper critical field, $H_{c2}(0)$ $\sim$ -0.7$T_c$(d$H_{c2}$/d$T$) is 3.7 T and 40.4 T for $H$$\parallel$ $c$ and $H$$\parallel$ $ab$ directions respectively. Also the respective Ginzburg-Landau (GL) coherence lengths, $\xi(0)$ = $(\phi_0/(2\pi H_{c2}(0)))^{1/2}$, are $\sim$ 94 \AA~and $\sim$ 28 \AA. Similar to the other studies on electron-doped cuprates \cite{Wang:2007, Cirillo:2009, Fournier:2003, Han:1992}, the irreversibility line, $H_{irr}(T)$ shows positive curvature for both directions and nicely fit to an empirical relation, $H_{irr}(T)$ = $H_{irr}(0)[1-(T/T_c)]^{\beta}$, where, $T_c$ = 23 K and $H_{irr}(0)$, $\beta$ are fitting parameters. The solid lines in $H_{irr}(T)$ show the fitted curve with parameters, $H_{irr}(0)$ = 3.8 $\pm$ 1 T, $\beta$ = 2.3 $\pm$ 0.1 for $H$$\parallel$$c$ and $H_{irr}(0)$ = 30.0 $\pm$ 3 T, $\beta$ = 2.6 $\pm$ 0.1 for $H$$\parallel$ $ab$ respectively. It is worth to mention that the values of the exponent $\beta$ are relatively higher than that found for YBaCuO, $\beta$$=$1.5 \cite{yeshurun:1988} and also larger than that found for the vortex-glass state \cite{Baruch:2010}. The anisotropy parameter, $\gamma_{H{irr}}$ = $H_{irr}$,$_{ab}$/$H_{irr}$,$_c$ is estimated as $\sim$ 8, which is close to the value estimated through slopes of $T_c(H)$-line. However, the anisotropy parameter obtained in the present study is twice smaller than the one reported in literature for this system \cite{Wang:2007} but nearly double compared to YBCO \cite{Welp:1989}. Contrary to the $H_{irr}(T)$ and $T_c(H)$, the $H_p(T)$ shows a negative curvature and well fitted using an empirical expression, $H_{p}(T)$ = $H_{p}(0)[1-(T/T_p)^{\mu}]$, where, $H_{p}(0)$, $T_p$ and $\mu$ are fitting parameters. The solid line for $H_p(T)$ in Fig. \ref{fig:HT}(b) shows the fitted curve with $H_{p}(0)$ = 1.64 $\pm$ 0.4 T, $T_p$ = 16.7 $\pm$ 0.2 K and $\mu$ = 1.5 $\pm$ 0.1. On the other hand, the $H_{on}$ (field associated to the onset of SMP in isothermal $M(H)$) shows an anomalous non-monotonic temperature dependence, which has not been observed previously in a superconductor. From $H_{c1}$ = 23 mT and $H_{c2}$(0) = 3.7 T for $H$$\parallel$ c-axis and the expression $H_{c2}$/$H_{c1}$ = 2$\kappa$$^2$ we estimate $\kappa$ $\approx$ 10 and from $H_{c1}$$\sim$$\phi$$_0$ln$\kappa$/4$\pi$$\lambda$$^2$ where $\lambda$ is the superconducting penetration depth \cite{tinkham}, and $H_{c1}$ = 2.4 mT for $H$$\parallel$ ab, we estimate $\lambda$(2K) $\sim$ 1250 \AA~for $H$$\parallel$ c-axis and $\lambda$(2K) $\sim$ 3890 \AA~for $H$$\parallel$ ab-planes. The value of $\lambda$ for $H$$\parallel$ c-axis is comparable to that obtained for YBaCuO ($T_c$=93K) $\lambda$(0) $\sim$~1460 \AA~and about half the value for the electron doped PrCeCuO4, where $\lambda$(0) $\sim$ 2790~\AA \cite{prozorov:2000}.

\section{Summary and Conclusion}

In summary, we studied the vortex dynamics in a single crystal of an electron doped cuprate superconductor Pr$_{0.87}$LaCe$_{0.13}$CuO$_4$, with $T_c$ $\sim$ 23 K. An unusual SMP  feature is observed in the initial branch of isothermal $M(H)$ for $H$$\parallel$$ab$-planes for $T$ $\leq$ 15 K. However, no such feature is observed for $H$$\parallel$$c$-axis direction. The origin of the SMP is explored using magnetic relaxation measurements at several fixed temperatures and magnetic fields. While the temperature and magnetic field dependence of the relaxation rate do not show any evidence of pinning crossover, a detailed analysis based on the activation pinning energy suggests that the SMP is associated to an elastic to plastic pinning crossover. Using Bean's critical state model, the critical current density is obtained from the isothermal $M(H)$ measurements for $H$$\parallel$$c$-axis and $H$$\parallel$$ab$-planes. The temperature dependence of critical current density, $J_c(T)$ for $H$$\parallel$$c$-axis does not follow the conventional pinning model based on variation in $T_c$ and mean free path. On the other hand, a peak is observed near 7 K in $J_c(T)$ for $H$$\parallel${ab}-planes. Such peak in $J_c(T)$ is likely to be related to a vortex-lattice dimensional crossover from a 3D Abrikosov to 2D Josephson vortices at low temperature, however its exact origin is yet to be found. $H$-$T$ phase diagrams for both crystal directions were constructed for different characteristic magnetic field and temperature values via making use of isothermal $M(H)$ and isofield $M(T)$ measurements. A relatively small anisotropy parameter, $\gamma$ $\sim$ 8-11 obtained in the zero temperature limit suggests the 3D nature of vortex lattice mainly for $H$$\parallel$$c$-axis.

\acknowledgements{}  
SSS was supported by CNPq, PVL acknowledges an undergraduated scientific fellowship from FAPERJ (Rio de Janeiro, Brazil, process E-26/201.915/2019), MPK acknowledges an undergraduated scientific fellowship from PIBIC-CNPq/UFRJ, LG was supported by FAPERJ (Process E-26/202.820/2018 and E-26/010/101136/2018) and also by CNPq. The work at IOP, CAS is supported by the National Natural Science Foundation of China (Nos.: 11374346, 11822411, 11961160699, 11874401, 11674406 and 11674372), the Strategic Priority Research Program (B) of the Chinese Academy of Sciences (CAS) (Nos.:  XDB07020300, XDB25000000), and the Youth Innovation Promotion Association of CAS (No.2016004).
\bibliographystyle{apsrev}

\begin{thebibliography}{0}
\expandafter\ifx\csname natexlab\endcsname\relax\def\natexlab#1{#1}\fi
\expandafter\ifx\csname bibnamefont\endcsname\relax
  \def\bibnamefont#1{#1}\fi
\expandafter\ifx\csname bibfnamefont\endcsname\relax
  \def\bibfnamefont#1{#1}\fi
\expandafter\ifx\csname citenamefont\endcsname\relax
  \def\citenamefont#1{#1}\fi
\expandafter\ifx\csname url\endcsname\relax
  \def\url#1{\texttt{#1}}\fi
\expandafter\ifx\csname urlprefix\endcsname\relax\def\urlprefix{URL }\fi
\providecommand{\bibinfo}[2]{#2}
\providecommand{\eprint}[2][]{\url{#2}}

\end{thebibliography}


\begin{thebibliography}{}

\bibitem{Melhem:2012} \textsl{High Temperature Superconductors (HTS) for Energy Applications}, Woodhead Publishing, (2012), edited by Ziad Melhem. 

\bibitem{Abulafia:1996} Y. Abulafia, A. Shaulov, Y. Wolfus, R. Prozorov, L. Burlachkov, Y. Yeshurun, D. Majer, E. Zeldov, H. W\"{u}hl, V. B. Geshkenbein, and V. M. Vinokur, Plastic Vortex Creep in YBa$_2$Cu$_3$O$_{7-x}$ Crystals, Phys. Rev. Lett. {\bf 77}, 1596 (1996).

\bibitem{Yang:1993} G. Yang, P. Shang, S. D. Sutton, I. P. Jones, J. S. Abell, and C. E. Gough, Competing Pinning Mechanisms in Bi$_2$Sr$_2$CaCu$_2$O$_y$ Single Crystals by Magnetic and Defect Structural Studies, Phys. Rev. B {\bf 48}, 4054 (1993).

\bibitem{Baruch:1999} B. Rosentein and A. Knigavko, Anisotropic Peak Effect due to Structural Phase Transition in the Vortex Lattice, Phys. Rev. Lett. {\bf 83}, 844 (1999).

\bibitem{Prozorov:2008} R. Prozorov, N. Ni, M. A. Tanatar, V. G. Kogan, R. T. Gordon, C. Martin, E. C. Blomberg, P. Prommapan, J. Q. Yan, S. L. Bud’ko, and P. C. Canfield, Vortex Phase Diagram of Ba(Fe$_{0.93}$Co$_{0.07}$)$_2$As$_2$ Single Crystals, Phys. Rev. B {\bf 78}, 224506 (2008).

\bibitem{Senatore:2008} C. Senatore, R. Flükiger, M. Cantoni, G. Wu, R. H. Liu, and X. H. Chen, Upper Critical Fields Well Above 100 T for the Superconductor SmFeAsO$_{0.85}$F$_{0.15}$ with $T_c$ = 46 K, Phys. Rev. B {\bf 78}, 054514 (2008).

\bibitem{Yang:2008} H. Yang, H. Luo, Z. Wang, and H.H. Wen, Fishtail Effect and the Vortex Phase Diagram of Single Crystal Ba$_{0.6}$K$_{0.4}$Fe$_2$As$_2$, Appl. Phys. Lett. {\bf 93}, 142506 (2008). 

\bibitem{SSugui:2010} S. Salem-Sugui, L. Ghivelder, A. D. Alvarenga, L. F. Cohen, K. A. Yates, K. Morrison, J. L. Pimentel, H. Luo, Z. Wang, and H-H Wen, Flux Dynamics Associated with the Second Magnetization Peak in the Iron Pnictide Ba$_{1-x}$K$_x$Fe$_2$As$_2$, Phys. Rev. B {\bf 82}, 054513 (2010).


\bibitem{Lortz:2007} R. Lortz, N. Musolino, Y. Wang, A. Junod, and N. Toyota, Origin of the Magnetization Peak Effect in the Nb$_3$Sn Superconductor, Phys. Rev. B {\bf 75}, 094503 (2007).

\bibitem{Das:2009} P. Das, C. V. Tomy, H. Takeya, S. Ramakrishnan and A. K. Grover, Peak Effect Phenomena, Surface Superconductivity and Paramagnetic Meissner Effect in a Spherical Single Crystal of Niobium, J. Phys.: Conf. Ser. {\bf 150}, 052041 (2009).

\bibitem{Kwok:1994} W. K. Kwok, J. A. Fendrich, C. J. van der Beek, and G. W. Crabtree, Peak Effect as a Precursor to Vortex Lattice Melting in Single Crystal YBa$_2$Cu$_3$O$_{7-\delta}$, Phys. Rev. Lett. {\bf 73}, 2614 (1994).

\bibitem{Chikumoto:1992} N. Chikumoto, M. Konczkowski, N. Motohira and A.P. Malozemoff, Flux-Creep Crossover and Relaxation Over Surface Barriers in Bi$_2$Sr$_2$CaCu$_2$O$_8$ Crystals, Phys. Rev. Lett. {\bf 69}, 1260 (1992).

\bibitem{feng1997} Y. Feng, L. Zhou, J. Wen and N. Koshizuka, Fishtail Effect and its Mechanism in PMP Y-Ho-Ba-Cu-O Superconductors, Physica C {\bf 282-287}, 2177 (1997).


\bibitem{Wang:2007} Y. Wang, C. Ren, L. Shan, S. Li, P. Dai and H-H. Wen, Peak Effect Due to Josephson Vortices in Superconducting Pr$_{0.88}$LaCe$_{0.12}$CuO$_{4−\delta}$ Single Crystals, Phys. Rev. B {\bf 75}, 134505 (2007).

\bibitem{Kamihara:2006} Y. Kamihara, H. Hiramatsu, M. Hirano, R. Kawamura, H. Yanagi, T. Kamiya and H. Hosono, Iron-Based Layered Superconductor: LaOFeP, J. Am. Chem. Soc. {\bf 128}, 10012 (2006).

\bibitem{Wilson:2006} S. D. Wilson, S. Li, P. Dai, W. Bao, J-H Chung, J. J. Kang, S-H Lee, S. Komiya, Y. Ando and Q. Si, Evolution of Low-Energy Spin Dynamics in the Electron-Doped High-Transition-Temperature Superconductor Pr$_{0.88}$LaCe$_{0.12}$CuO$_{4-\delta}$, Phys. Rev. B {\bf 74}, 144514 (2006).

\bibitem{Niestemski:2007} F. C. Niestemski, S. Kunwar, S. Zhou, S. Li, J. Ding, Z. Wang, P. Dai and V. Madhavan, A Distinct Bosonic Mode in an Electron-Doped High-Transition-Temperature Superconductor, Nature {\bf 450}, 1058 (2007).

\bibitem{Kang:2007} H. J. Kang, P. Dai, B. J. Campbell, P. J. Chupas, S. Rosenkranz, P. S. Lee, Q. Huang, S. Li, S. Komiya and Y. Ando, Microscopic Annealing Process and its Impact on Superconductivity in $T^{'}$-Structure Electron-Doped Copper Oxides, Nature Materials {\bf 6}, 224 (2007).

\bibitem{Wilson:2007} S. D. Wilson, S. Li, J. Zhao, G. Mu, H-H Wen, J. W. Lyin, P. G. Freeman, L-P Regnault, K. Habicht and P. Dai, Quantum Spin Correlations Through the Superconducting-to-Normal Phase Transition in Electron-Doped Superconducting Pr$_{0.88}$LaCe$_{0.12}$CuO$_{4-\delta}$, PNAS, {\bf 104}, 15259 (2007).

\bibitem{Li:2008} S. Li, S. Chi, J. Zhao, H-H Wen, M. B. stone, J. W. Lyin and P. Dai, Impact of Oxygen Annealing on the Heat Capacity and Magnetic Resonance of Superconducting Pr$_{0.88}$LaCe$_{0.12}$CuO$_{4-\delta}$. Phys. Rev. B {\bf 78}, 014520 (2008).

\bibitem{Fujita:2008} M. Fujita, M. Matsuda, S.-H. Lee, M. Nakagawa, and K. Yamada, Low-Energy Spin Fluctuations in the Ground States of Electron-Doped Pr$_{1−x}$LaCe$_x$CuO$_{4+\delta}$ Cuprate Superconductors, Phys. Rev. Lett.{\bf 101}, 107003 (2008).

\bibitem{Wang:2009} Y-L Wang, Y. Huang, L. Shan, S. L. Li, P. Dai, C. Ren and H-H Wen, Annealing Effect on the Electron-Doped Superconductor Pr$_{0.88}$LaCe$_{0.12}$CuO$_{4\pm \delta}$, Phys. Rev. B {\bf 80}, 094513 (2009).

\bibitem{Adachi:2013} T. Adachi, Y. Mori, A. Takahashi, M. Kato, T. Nishizaki, T. Sasaki, N. Kobayashi, and Y. Koike, Evolution of the Electronic State through the Reduction Annealing in Electron-Doped Pr$_{1.3-x}$La$_{0.7}$Ce$_x$CuO$_{4+\delta}$ ($x$=0.10) Single Crystals: Antiferromagnetism, Kondo Effect, and Superconductivity, J. Phys. Soc. Jpn. {\bf 82}, 063713 (2013).

\bibitem{Horio:2016} M. Horio, T. Adachi, Y. Mori, A. Takahashi, T. Yoshida, H. Suzuki, L. C. C. Ambolode II, K. Okazaki, K. Ono, H. Kumigashira, H. Anzai, M. Arita, H. Namatame, M. Taniguchi, D. Ootsuki, K. Sawada, M. Takahashi, T. Mizokawa, Y. Koike, and A. Fujimori, Suppression of the Antiferromagnetic Pseudogap in the Electron-Doped High-Temperature Superconductor by Protect Annealing, Nat. Commun. {\bf 7}, 10567 (2016).

\bibitem{Adachi:2016} T. Adachi, A. Takahashi, K. M. Suzuki, M. A. Baqiya, T. Konno, T. Takamatsu, M. Kato, I. Watanabe, A. Koda, M. Miyazaki, R. Kadono, and Y. Koike, Strong Electron Correlation behind the Superconductivity in Ce-Free and Ce-Underdoped High-$T_c$ $T^{'}$-Cuprates, J. Phys. Soc. Jpn. {\bf 85}, 114716 (2016).

\bibitem{Yamamoto:2016} M. Yamamoto, Y. Kohori, H. Fukazawa, A. Takahashi, T. Ohgi, T. Adachi, and Y. Koike, Existence of Large Antiferromagnetic Spin Fluctuations in Ce-Doped $T^’$-Cuprate Superconductors, J. Phys. Soc. Jpn. {\bf 85}, 024708 (2016).

\bibitem{Adachi:2017} T. Adachi, T. Kawamata, and Y. Koike, Novel Electronic State and Superconductivity in the Electron-Doped High-$T_c$ $T^{'}$-Superconductors, Condens. Matter {\bf 2}, 23 (2017).

\bibitem{Baqiya:2019} M. A. Baqiya,T. Adachi, A. Takahashi, T. Konno, T. Ohgi, I. Watanabe, and Y. Koike, Muon-Spin Relaxation Study of the Spin Correlations in the Overdoped Regime of Electron-Doped High-$T_c$ Cuprate Superconductors, Phys. Rev. B {\bf 100}, 064514 (2019).


\bibitem{Lambacher:2010} M. Lambacher, T. Helm, M. Kartsovnik, and A. Erb, Advances in single crystal growth and annealing treatment of electron-doped HTSC, Eur. Phys. J. Special Topics {\bf 188}, 61 (2010).

\bibitem{veal} B. W. Veal, A. P. Paulikas, H. You, H. Shi, Y. Fang, and J. W. Downey, Observation of temperature-dependent site disorder in YBa$_2$Cu$_3$0$_{7-\delta)}$ below $^\circ$C, Phys. Rev. B {\bf 42}, 6305 (1990).

\bibitem{deak} J. Deak, Lifang Hou, P. Metcalf, and M. McElfresh,  Dimensional crossover field as a function of oxygen stoichiometry in YBa$_2$Cu$_3$0$_{7-\delta}$ thin films Phys. Rev. B {\bf 51}, 705 (1994).


\bibitem{Roeser:2009} H. P. Roeser, F. M.Huber, M. F. von Schoenermark, A. S. Nikoghosyan , F. Hetﬂeisch, M. Stepper and A. Moritz, Doping patterns in N-type high temperature superconductors PLCCO and NCCO, Acta Astronautica {\bf 65}, 289 (2009).

\bibitem{Mints:1981} R. G. Mints and A. L. Rakhmanov, Critical State Stability in Type-II Superconductors and Superconducting-Normal-Metal Composites, Rev. Mod. Phys. {\bf 53}, 551 (1981). 

\bibitem{Sundar:2015} Shyam Sundar, M. K. Chattopadhyay, L. S. Sharath Chandra and S. B. Roy, Magnetic Irreversibility and Pinning Force Density in the Mo$_{100-x}$Re$_x$ Alloy Superconductors, Physica C {\bf 519}, 13 (2015).

\bibitem{Lee:2015} H. B. Lee, G. C. Kim, Y. C. Kim, D. Ahmad and Y. S. Kwon, Flux Jump Behaviors and Mechanism of MgB$^2$ Synthesized by the Non-special Atmosphere Synthesis, J. Supercond. Nov. Magn. {\bf 28}, 2663 (2015).

\bibitem{Li:2013} J. Li, Y. Guo, S. Zhang, Y. Tsujimoto, X. Wang, C. I. Sathish, S. Yu, Y. Sun, W. Yi, K. Yamaura, E. Takayama-Muromachi, Y. Shirako, M. Akaogi and L.E. De Long, Quasi-periodic Magnetic Fux Jumps in the Superconducting State of Ba$_{0.5}$K$_{0.5}$Fe$_{1.9}$M$_{0.1}$As$_2$ ( M = Fe, Co, Ni, Cu, and Zn), Physica C {\bf 495}, 192 (2013).

\bibitem{Garber:1993} A. Gerber, Z. Tarnawski, J. J. M. Franse, J. N. Li and A. A. Menovsky, Flux Jump in High $T_c$ Single Crystals, Applied Superconductivity {\bf 1}, 961 (1993).

\bibitem{Maley:1990} M. P. Maley, J. O. Willis, H. Lessure, and M. E. McHenry, Dependence of Flux-Creep Activation Energy Upon Current Density in Grain-Aligned YBa$_2$Cu$_3$O$_{7-x}$, Phys. Rev. B {\bf 42}, 2639(R) (1990).

\bibitem{McHenry:1991} M. E. McHenry, S. Simizu, H. Lessure, M. P. Maley, J. Y. Coulter, I. Tanaka, and H. Kojima, Dependence of the Flux-Creep Activation Energy on the Magnetization Current for a La$_{1.86}$Sr$_{0.14}$CuO$_4$ Single Crystal, Phys. Rev. B {\bf 44}, 7614 (1991).

\bibitem{Feigel'man:1989} M. V. Feigel'man, V. B. Geshkenbein, A. I. Larkin, and V. M. Vinokur, Theory of Collective Flux Creep, Phys. Rev. Lett. {\bf 63}, 2303 (1989).

\bibitem{Sundar1:2017} S. Sundar, S. Salem-Sugui Jr., H. S. Amorim, Hai-Hu Wen, K. A. Yates, L. F. Cohen, and L. Ghivelder, Plastic Pinning Replaces Collective Pinning as the Second Magnetization Peak Disappears in the Pnictide Superconductor Ba$_{0.75}$K$_{0.25}$Fe$_2$As$_2$, Phys. Rev. B {\bf 95}, 134509 (2017).

\bibitem{Sundar2:2017} S. Sundar, J. Mosqueira, A. D. Alvarenga, D. S\'{o}\~{n}ora, A. S. Sefat, and S. Salem-Sugui Jr., Study of the Second Magnetization Peak and the Pinning Behaviour in Ba(Fe$_{0.935}$Co$_{0.065}$)$_2$As$_2$ Pnictide Superconductor, Supercond. Sci. Technol. {\bf 30}, 125007 (2017).

\bibitem{Sundar:2019} S. Sundar, S. Salem-Sugui Jr., E. Lovell, A. Vanstone, L. F. Cohen, D. Gong, R. Zhang, X. Lu, H. Luo and L. Ghivelder, Doping Dependence of the Second Magnetization Peak, Critical Current Density, and Pinning Mechanism in BaFe$_{2-x}$Ni$_x$As$_2$ Pnictide Superconductors, ACS Appl. Electron. Mater. {\bf 1}, 179, (2019).

\bibitem{Tachiki:1989} M. Tachiki and S. Takahashi, Strong Vortex Pinning Intrinsic in High-$T_c$ Oxide Superconductors, Solid State Communication {\bf 70}, 291 (1989).

\bibitem{Bean:1964} C. P. Bean, Magnetization of High-Field Superconductors. Rev. Mod. Phys. {\bf 36}, 31 (1964); A. Umezawa, G. W. Crabtree, J. Z. Liu, H. W. Weber, W. K. Kwok, L. H. Nunez, T. J. Moran, C. H. Sowers and H. Claus, Enhanced Critical Magnetization Currents Due to Fast Neutron Irradiation in Single-Crystal YBa$_2$Cu$_3$O$_{7-\delta}$, Phys. Rev. B {\bf 36}, 7151 (1987).

\bibitem{gennep:2020} D. V. Gennep, A. Hassan, H. Luo and M. Abdel-Hafiez, Sharp peak of the critical current density in BaFe$_{2-x}$Ni$_x$As$_2$ at optimal composition, Phys. Rev. B {\bf 101}, 235163 (2020).

\bibitem{Griessen:1994} R. Griessen, H-H Wen, A. J. J. van Dalen, B. Dam, J. Rector, H. G. Schnack, S. Libbrecht, E. Osquiguil, and Y. Bruynseraede, Evidence for Mean Free Path Fluctuation Induced Pinning in YBa$_2$Cu$_3$O$_7$ and YBa$_2$Cu$_4$O$_8$ films. Phys. Rev. Lett. {\bf 72}, 1910 (1994).

\bibitem{Hosseinzadeh:2019} M. Hosseinzadeh, S. R. ghorbani and H. Arabi, On the Determination of Pinning Mechanisms and Regimes in Type-II Superconductors with Weak Thermal Fluctuations, J Supercond Nov Magn {\bf 33}, 971 (2020).

\bibitem{Hu:1998} Xiao Hu and Masashi Tachiki, Structure and Phase Transition of Josephson Vortices in Anisotropic High-$T_c$ Superconductors, Phys. Rev. Lett. {\bf 80}, 4044 (1998).

\bibitem{Fehresnbacher:1992}Roland Fehrenbacher, Vadim B.Geshkenbein, Gianni Blatter, Pinning phenomena and critical currents in disordered long Josephson junctions, Phys. Rev. B {\bf 45}, 5450 (1992).

\bibitem{Gurevich:1992}A. Gurevich, Nonlocal Josephson electrodynamics and pinning in superconductors, Phys. Rev. B {\bf 46}R, 3187 (1992)

\bibitem{Baruch:2001} B. Rosenstein, B. Ya. Shapiro, R. Prozorov, A. Shaulov, and Y. Yeshurun, Fluctuations in Single-Crystal YBa$_2$Cu$_3$O$_{6.5}$: Evidence for Crossover from Two-Dimensional to Three-Dimensional Behavior, Phys. Rev. B {\bf 63}, 134501 (2001).

\bibitem{Moll:2013} P. J. W. Moll, L. Balicas, V. Geshkenbein, G. Blatter, J. Karpinski, N. D. Zhigadlo, and B. Batlogg, Transition from Slow Abrikosov to Fast Moving Josephson Vortices in Iron Pnictide Superconductors, Nature Materials {\bf 12}, 134 (2013).

\bibitem{Egilmez:2019} M. Egilmez, I. Isaac, A.S. Alnaser, Z. Bukowski, J. Karpinski, K. H. Chow  and J. Jung, Instabilities of the Vortex Lattice and the Peak Effect in Single Crystal YBa$_2$Cu$_4$O$_8$, Condens. Matter {\bf 4}, 74 (2019).

\bibitem{putilov:2019} A. V. Putilov, C. D. Giorgio, V. L. Vadimov, D. J. Trainer, E. M. Lechner, J. L. Curtis, M. Abdel-Hafiez, O. S. Volvoka, A. N. Vasiliev, D. A. Chareev, G. Karapetrov, A. E. Koshelev, A. Y. Aladyshkin, A. S. Melnikov, and M. Iavarone, Vortex core properties and vortex-lattice transformation in FeSe, Phys. Rev. B {\bf 99}, 144514 (2019).

\bibitem{Welp:1989} U. Welp, W. K. Kwok, G. W. Crabtree, K. G. Vandervoort and J. Z. Liu, Magnetic Measurements of the Upper Critical Field of YBa$_2$Cu$_3$O$_{7-\delta}$ Single Crystals, Phys. Rev. Lett. {\bf 62}, 1908 (1989).

\bibitem{Cirillo:2009} C. Cirillo, A. Guarino, A. Nigro, C. Attanasio, Critical Currents and Pinning Forces in Nd$_{2-x}$Ce$_x$CuO$_{4-\delta}$ thin films, Phys. Rev. B {\bf 79}, 144524 (2009).

\bibitem{Fournier:2003} P. Fournier and R. L. Greene, Doping Dependence of the Upper Critical Field of Electron-Doped Pr$_{2-x}$Ce$_x$CuO$_4$ Thin Films, Phys. Rev. B {\bf 68}, 094507 (2003).

\bibitem{Han:1992} S. H. Han, C. C. Almasan, M. C. de Andrade, Y. Dalichaouch, and M. B. Maple, Determination of the Upper Critical Field of the Electron-Doped Superconductor Sm$_{1.85}$Ce$_{0.15}$CuO$_{4-y}$ from Resistive Fluctuations, Phys. Rev. B {\bf 46}, 14290 (1992).

\bibitem{yeshurun:1988}Y. Yeshurun and A.P. Malozemoff, Giant Flux Creep and Irreversibility in an Y-Ba-Cu-O Crystal: An Alternative to the Superconducting-Glass, Model Phys. Rev. Lett. {\bf 60}, 2202 (1988).

\bibitem{Baruch:2010}Baruch Rosenstein and Dingping Li, Ginzburg-Landau theory of type II superconductors in magnetic field, Rev. Mod. Phys. {\bf 82}, 109 (2010).

\bibitem{tinkham}M. Tinkham, Introduction to Superconductivity, 2nd Edition McGraw-Hill, Inc., New York (1996).

\bibitem{prozorov:2000}R. Prozorov, R. W. Giannetta, A. Carrington, P. Fournier, R. L. Greene, P. Guptasarma, D. G. Hinks, and A. R. Banks, Measurements of the absolute value of the penetration depth in high- $T_c$ superconductors using a low- $T_c$ superconductive coating,  Applied Physics Letters {\bf 77}, 4202 (2000).

\end{thebibliography}

\end{document}